\documentclass[twocolumn]{openjournal}
\pdfoutput=1

\usepackage{graphicx} 
\usepackage{amssymb}    
\usepackage{amsmath}
\usepackage{xcolor}
\usepackage{booktabs}
\usepackage{dblfloatfix}
\usepackage{placeins}
\usepackage{graphicx}
\usepackage{tikz}

\usepackage{pgfplotstable}
\usepackage{threeparttable}

\usepackage{booktabs}
\pgfplotsset{compat=1.18}

\usepackage{natbib}
\usepackage{hyperref}

\begin{document}

\shorttitle{Temperature effects on dust aggregates}
\shortauthors{Kolanz L., Lazzati D., Guidos J.}

\title{Effect of temperature on the structure of porous dust aggregates formed by coagulation}

\author{Lucas Kolanz, Davide Lazzati, Job Guidos}

\affil{Department of Physics, Oregon State University, 301
  Weniger Hall, Corvallis, OR 97331, USA} 

\begin{abstract}

The source of high redshift dust is currently under debate. One possibility  are the ejecta of pair-instability and core collapse supernovae. However, it is uncertain how much newly formed dust can survive the supernova reverse shock and be injected into the interstellar medium. We anticipate the structure of the pre-shocked dust to affect how much of it survives. Yet, the structure of dust formed in supernova is not well understood. We present three-dimensional soft-sphere, dust coagulation simulations, using sequential collisions, aimed at studying the impact of temperature and monomer size distribution on the structure of growing dust aggregates. Due to the qualitative nature of the concept of structure, there are many ways to define and quantify it, especially for an irregular aggregate. Thus, we test eight metrics commonly used in the literature in order to compare the aggregate properties as well as the strengths and weaknesses of the metrics themselves. Our findings show that higher temperatures result in denser, more compact structures for all metrics tested. Additionally, we find that structures that coagulate from a distribution of monomer sizes are denser and more compact than structures formed from identical monomers under similar conditions. The latter finding, however, is true for all of the metrics except for the average number of contact points, which has proven to be the least reliable of the eight considered metrics.
\end{abstract}

\section{Introduction} \label{sec:intro}

Cosmic dust is not only the building block of asteroids, comets, and planetesimals, but also accounts for significant fractions of the total measured radiation of some galaxies by absorbing stellar radiation (up to 50\% in some cases) and re-emitting photons at longer wavelengths \citep{Calzetti2001}. How dust interacts with radiation depends, in part, on its structure and other physical properties. The structure of dust in various cosmic environments has been estimated using N-body dust aggregation simulations \citep{Tanaka2012}, various analytical models (such as EMT-Mie theory and the Discrete Dipole Approximation)\citep{Ysard2018}, experiments \citep{Wurm2000}, and dust collected from the deep sea, space, and stratosphere \citep{Brownlee1985}.

There are two main known sources of cosmic dust. The first, asymptotic giant branch (AGB) stars, go through a phase of high stellar winds and mass loss, up to $10^{-4} M_{\odot} yr^{-1}$ \citep{Bowen1991}. Dust can then form in the AGB star's cooling ejecta. A single AGB star can produce between $10^{-5}$ and $10^{-2} M_{\odot}$ of dust \citep{Ventura2012}.
However, AGB stars are not traditionally thought to be possible dust sources in the early universe ($z \gtrsim 6$). This is because it takes time for these stars to evolve enough to produce an environment suitable for dust to form. This discrepancy was recently magnified by the James Webb Space Telescope discovery of galaxies at $z \approx 7$ with dust masses around $10^{8} M_{\odot}$ \citep{Witstok2023}. In this case, observed dust is often attributed to the second main source of cosmic dust, supernovae (SNe).  

Dust formation in SNe occurs only once expansion produces temperatures and densities which are favorable for the gasses present to nucleate. SNe dust formation is somewhat unique in astrophysics, as it varies over a timescale of decades instead of the usual cosmological timescales of hundreds of thousands to billions of years. However, in order to reach the interstellar medium (ISM) the newly formed dust still needs to survive the supernova reverse shock (e.g., \citealt{Nozawa2007}). 

Depending on the density of the ISM around the supernova remnant (SNR), the size of the progenitor star, the type of dust, and the physical characteristics of the dust particles at the time of the reverse shock, 50-98\% of the newly formed dust can be destroyed in this event \citep{Bianchi2007,Slavin2020}. Recently, \cite{Kirchschlager2024} has also shown that dust destruction is highly dependent on the temporal evolution of the SN ejecta. However, to our knowledge, no studies on reverse shock dust destruction take into account dust aggregates, and only consider dust as consisting of single monomers. 

There have been discrete element simulations of dust aggregation done in the past, but they have been primarily through the lens of aggregation in a protoplanetary disk \citep{Dominik1997,Wada2007,Wada2008,Wada2009,Wada2011,Wada2013,Suyama2008,Okuzumi2009,Paszun2009,Seizinger2012,Seizinger2013,Hasegawa2021,Hasegawa2023} and thus the structure of growing aggregates was of less significance than properties such as collision outcomes. In addition, most work has been carried out with aggregates formed with monomers of equal size using sequential sticking codes \citep{Wada2007,Wada2008,Wada2009,Wada2011,Wada2013,Okuzumi2009,Paszun2009,Seizinger2012,Seizinger2013,Hasegawa2021,Hasegawa2023}, that simply attach a new monomer to the growing aggregate, neglecting any local or global restructuring as a result of the energy deposition.

Our goal with this research is to gain a better understanding of the possible structure of aggregates formed in an SNR environment before they interact with the reverse shock. Specifically, we simulate the formation of dust aggregates using a soft-sphere discrete element method code called DECCO (Discrete Element Cosmic COllision, \citealt{Guidos2025}). 

This manuscript is organized as follows: in section \ref{sec:methods} we describe our simulation code and physical parameters, the aggregate growth scheme, and metrics of quantifying aggregate structure. In section \ref{sec:results} we describe the results of our simulations. Section \ref{sec:summary} summarizes and discusses our results and their implications. In the appendix we compare and discuss the porosity metrics used in this paper and also provide a table of average geometric cross sections for our aggregates.    

Throughout this work, we use the term ``monomer'' to refer to an individual, spherical, solid dust particle. A monomer is the smallest unit treated in our simulations and has a radius of the order $\sim 0.1\,\mu\mathrm{m}$. The term ``aggregate'' denotes a bound collection of monomers held together by attractive forces. When referring to ``lognormal aggregates'' or ``constant aggregates'', we mean that the radii of the monomers within the aggregate follow a lognormal or constant distribution, respectively.

\begin{figure*}
    \centering
    \includegraphics[width=0.98\textwidth]{ColoredAggComp_constrelax.png}

    \vspace{0.5\baselineskip}

    \includegraphics[width=0.98\textwidth]{ColoredAggComp_lognormrelax.png}

    \caption{Example aggregates of sizes $N = 30$, $100$, and $300$ monomers formed at temperatures $3$, $10$, $30$, $100$, $300$, and $1000$~K. Note that for all sizes, but especially for larger sizes, higher temperatures produce more compact aggregates. The top panel shows aggregates with a constant monomer-size distribution, while the bottom panel shows aggregates with a lognormal monomer-size distribution.}
    \label{fig:agg_comp}
\end{figure*}

\begin{figure*}
    \centering
    \includegraphics[width=\textwidth,height=0.90\textheight,keepaspectratio]{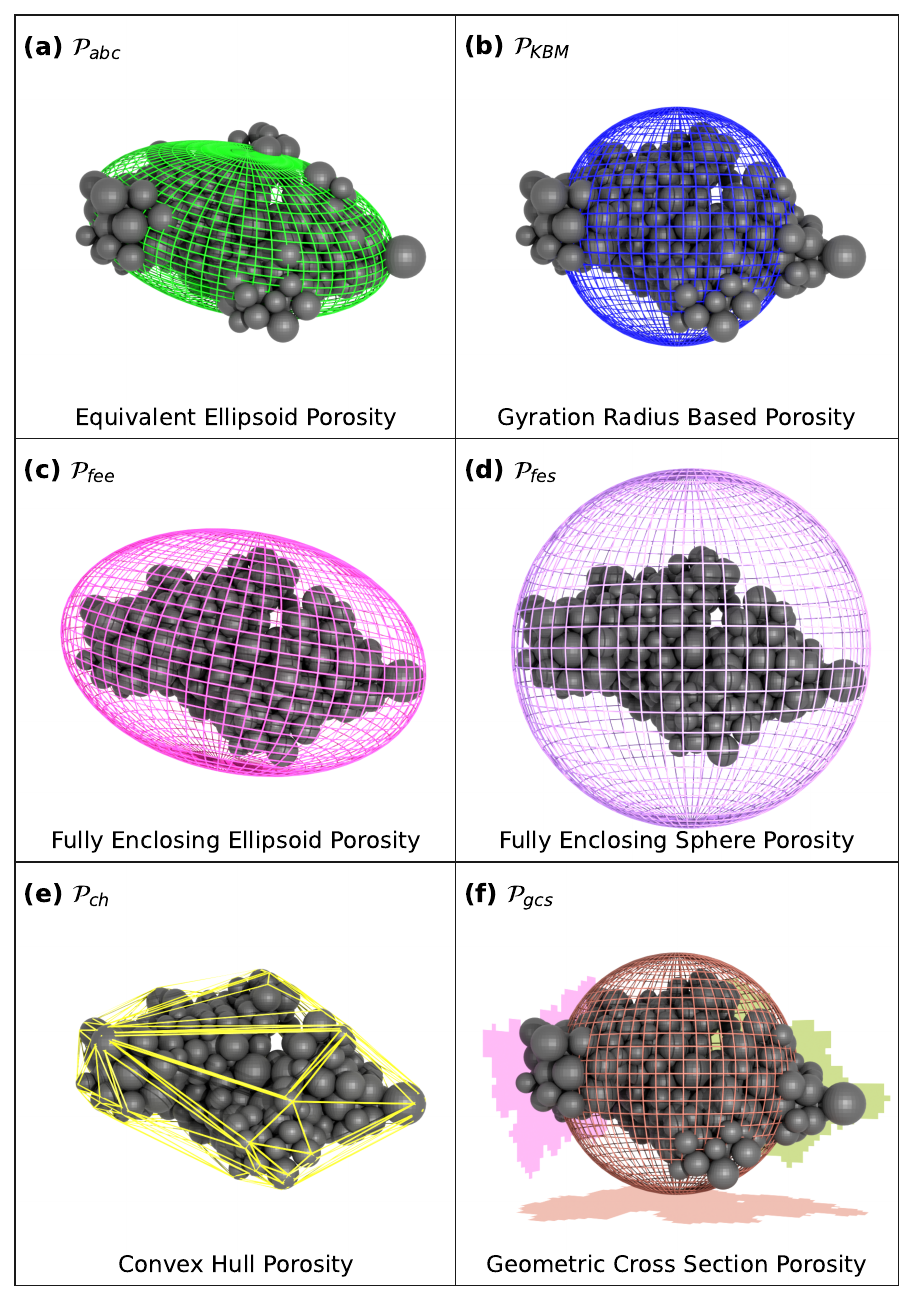}
    
    \caption{Visualizations of the porosities used in this paper. The hulls represent the volume occupied by the aggregate as defined by each porosity metric. Note that panel (e) shows a convex hull generated using 32 points per sphere for visualization purposes (as opposed to 64 used in calculations). Panel (f) shows the geometric cross section for three different directions, as well as the resulting sphere representing the volume the aggregate occupies. The grid shown in panel (f) is much larger than the one used for actual calculations.}
 \label{fig:porosityVisualization}
\end{figure*}

\section{Methods}\label{sec:methods}

\subsection{Numerical Simulation}
\label{subsec:NumSim}

In this study we ran three-dimensional, soft-sphere, discrete element dust coagulation simulations. We have carried out simulations with two monomer size distributions, constant and lognormal. In the constant monomer size distribution, individual monomers in our simulations are spherical with a radius of 0.1 $\mu m$. In the lognormal case, monomers are spherical, but are given a radius chosen from a lognormal distribution. The lognormal distribution is centered such that on average aggregates with the same number of monomers have the same mass for both monomer size distributions. The standard deviation of the lognormal distribution is set to $\sigma=0.2$, and simulated monomers range in size from 0.042 to 0.23 $\mu$m. We did not impose a direct restriction on the range of monomer sizes, however, in some cases a monomer was so big it would introduce numeric instabilities. In this case, the simulation would be rerun with another randomly selected monomer size. Monomers have a density of 2.25 g/cm$^{3}$ (consistent with amorphous carbon), and the aggregates grown range in radius (measured as the furthest monomer from the center of mass of the aggregate) from 0.35 to 1.5 $\mu$m.

The coefficient of friction of graphite varies greatly with ambient temperature, pressure, and gas species. For the coefficient of sliding friction we chose a value on the low end  $\mu_k=0.1$ \citep{morstein2022}. We also set the coefficient of rolling friction to a value on the low end $\mu_r=10^{-5}$. 

The code used in this study is an improved version of the DECCO\footnote{https://github.com/LucasKolanz/DECCO} (Discrete Element Cosmic COllision \citealt{Guidos2025}) code. In this version we changed the attractive force between two monomers from gravitational to Van der Waals forces using the equation \citep{Hamaker37}:

\begin{equation}
    F_{VDW}=-\frac{A}{6}\frac{64 r_{1}^{3} r_{2}^{3} z}{(z^{2}-(r_{1}+r_{2})^{2})^{2}(z^{2}-(r_{1}-r_{2})^{2})^{2}}
    \label{eq:VDWforce}
\end{equation}

\noindent where $A$ is the so-called Hamaker constant, $r_1$ and $r_2$ are the radii of the two monomers. The quantity $z$ represents the distance between the centers of the two monomers when the monomers are at a sufficient distance from each other. In practice $z$ is calculated as:

\begin{equation}
    z=r_1+r_2+\max(h,h_{\min})
    \label{eq:ctocdist}
\end{equation}

\noindent where $h$ is the distance between the surfaces of the two monomers. Thus, $h=0$ when monomers come into contact and $h<0$ when monomers are pushing against each other (as in other soft sphere codes). To prevent the divergency this causes in Equation~\eqref{eq:VDWforce}, a minimum value of $h$ ($h_{min}$) is set when $h<h_{min}$ in this equation such that $z$ is never allowed to be too close to the sum of the particle radii. This study sets $h_{min}=5\times10^{-6}$ cm (0.05 $\mu$m) for both the lognormal and constant monomer size distribution runs. 

The change from gravity to van der Waals forces was necessary since the original DECCO was developed for large constituent particles of asteroids, not small dust particles. Alternative analytic models for describing cohesive forces have been developed and studied, as we discuss below. Gravity has been neglected in this study as the force is negligible at the scale of dust. 
Take, for example, the case of two amorphous carbon dust grains 0.1 $\mu$m in radius, each with a mass of 9.42$\times 10^{-15}$ g, 0.1 $\mu$m apart. The Van der Waals force is over 10$^{19}$ times stronger than the force of gravity between them. 

Energy is dissipated in DECCO through friction forces, as well as through a constant coefficient of restitution. Since repulsive forces in DECCO are treated as spring forces, the coefficient of restitution is calculated as:

\begin{equation}
    \epsilon=\sqrt{k_{out}/k_{in}}
    \label{eq:COR}
\end{equation}

\noindent where $k_{in}$ is the spring constant during compression, $k_{out}$ is the spring constant during decompression, and $k_{in} > k_{out}$. In practice, $k_{in}$ is set such that there will be multiple time steps associated with any collision. Then, $k_{out}$ is set based on Equation~\eqref{eq:COR}. For a more in depth look at how the code handles this, see \cite{Guidos2025}. The benefit of this is a scale-free constant of restitution, useful for simulating many, different sized monomers, and is applicable from dust to asteroid/planetesimal scales.

\begin{figure*}
    \centering
    \includegraphics[width=\textwidth,height=0.93\textheight,keepaspectratio]{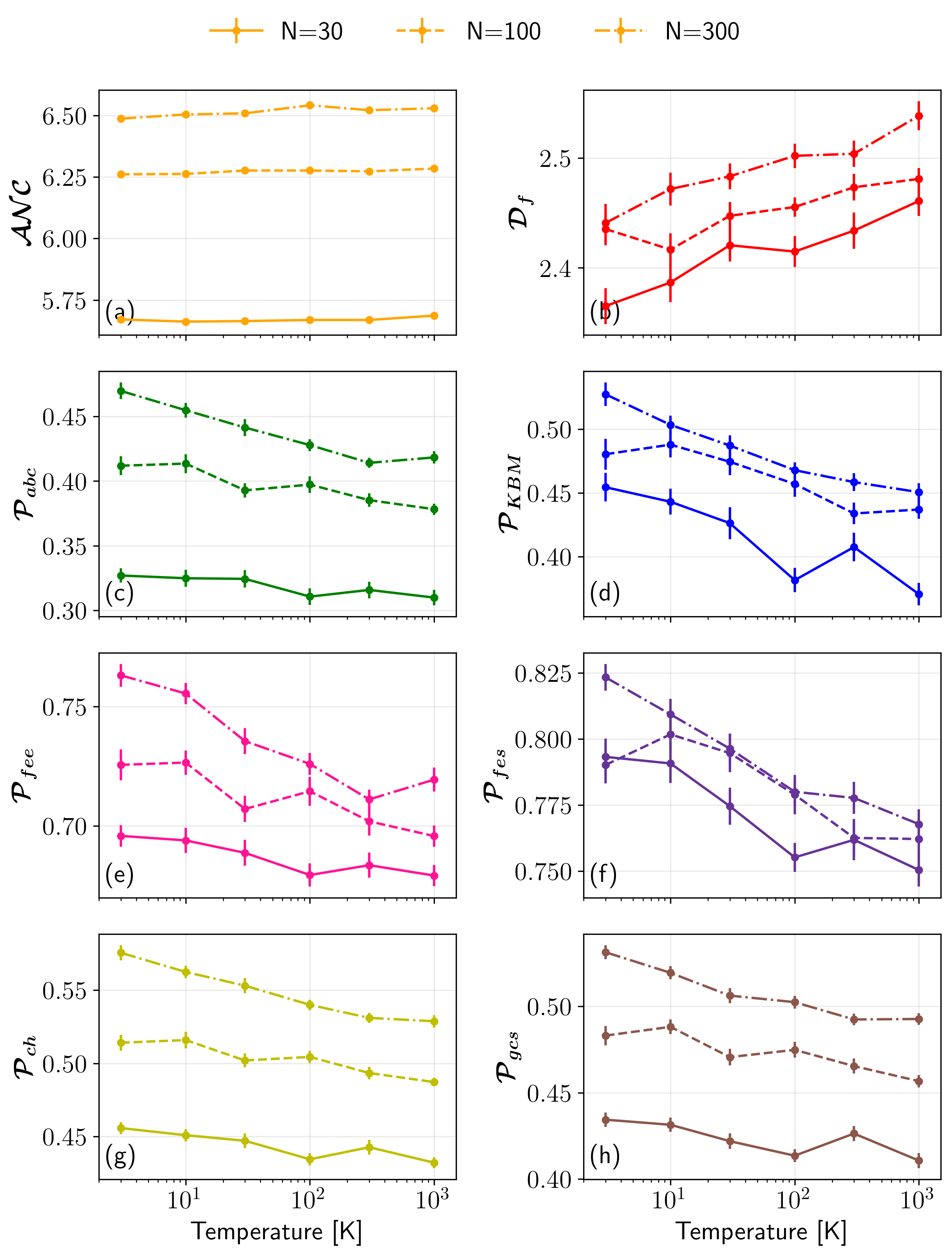}

    \caption{Structure metrics for the constant monomer-size aggregates at various temperatures and for various sizes. In order from (a)-(h) the panels show the average number of contacts, fractal dimension, $\mathcal{P}_{abc}$, $\mathcal{P}_{KBM}$, $\mathcal{P}_{fee}$, $\mathcal{P}_{fes}$, $\mathcal{P}_{ch}$, $\mathcal{P}_{gcs}$. Values for the same size aggregate at different temperatures are joined with a line (see the legend on top of panels (a) and (b)). Each symbol shows the average of 30 statistically independent realizations. The error bars represent the standard error of these sub-samples.} 
        \label{fig:const_measures}
\end{figure*}

\begin{figure*}
    \includegraphics[width=\textwidth,height=0.93\textheight,keepaspectratio]{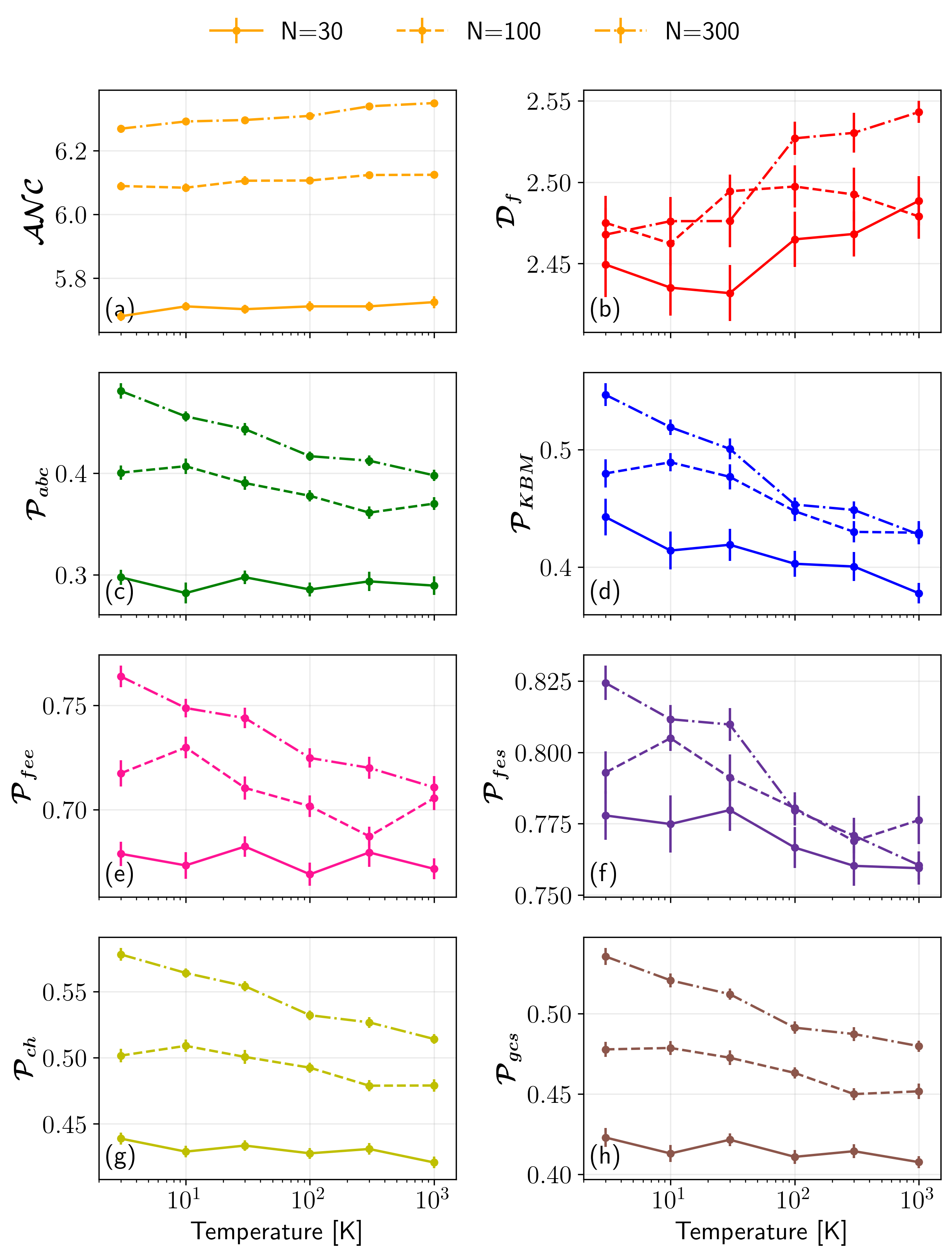}
    
    \caption{Same as Figure~\ref{fig:const_measures} but for aggregates with a lognormal monomer-size distribution.} 
    \label{fig:lognorm_measures}
\end{figure*}

\begin{figure*}
 \begin{center}
\parbox{1.0\textwidth}{
    \includegraphics[width=0.99\textwidth]{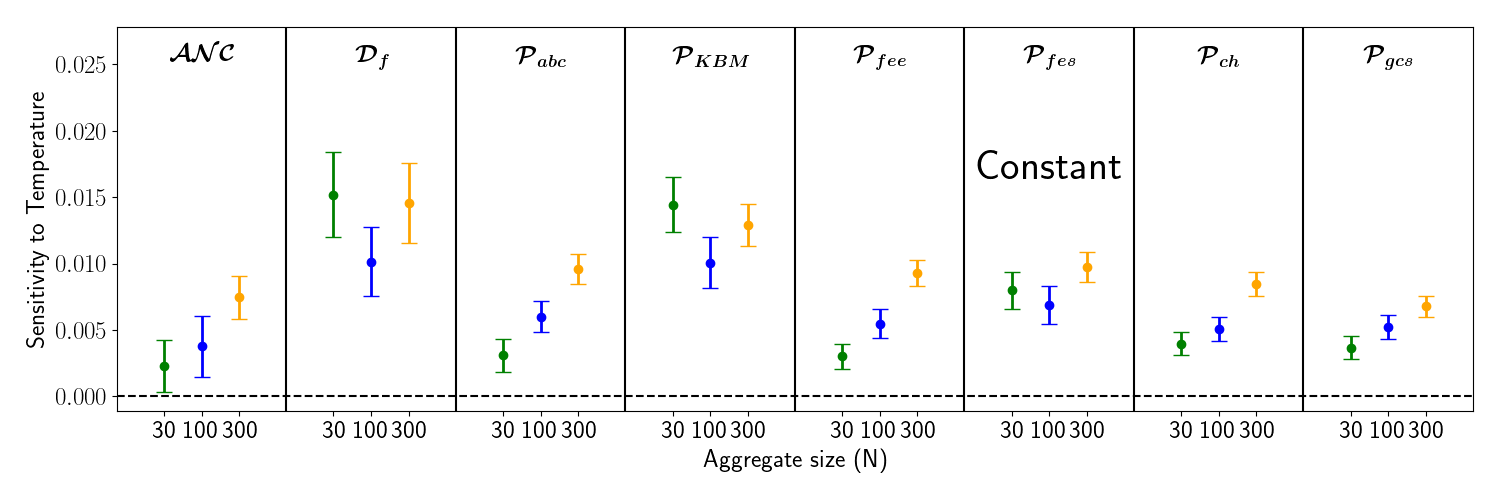}}
\parbox{1.0\textwidth}{
    \includegraphics[width=0.99\textwidth]{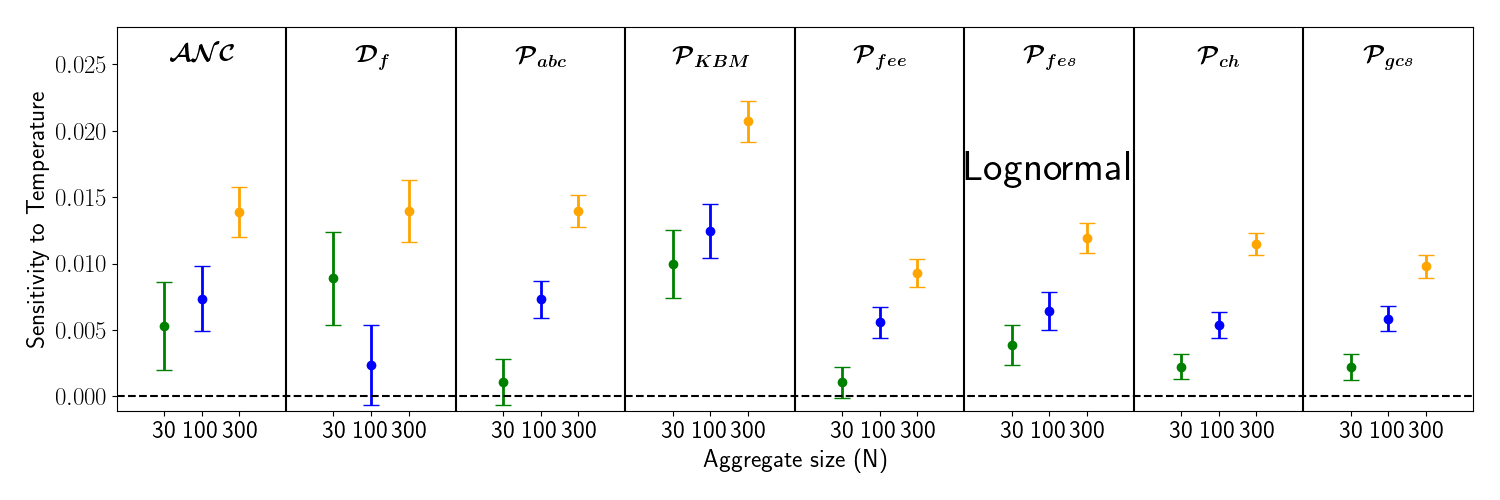}}
\end{center}
    \caption{Absolute value of the slope of a line fit to the data from Figures~\ref{fig:const_measures} and \ref{fig:lognorm_measures} as a way to quantify the sensitivity of an aggregate's structure to temperature, for a constant (top panel) and lognormal (bottom panel) monomer radii distribution. Vertical error bars represent uncertainty in the slope. In almost all cases a statistically significant temperature dependence is identified.}
    \label{fig:method_comp}
\end{figure*}

\begin{figure*}
    
\includegraphics[width=\textwidth,height=0.93\textheight,
keepaspectratio]{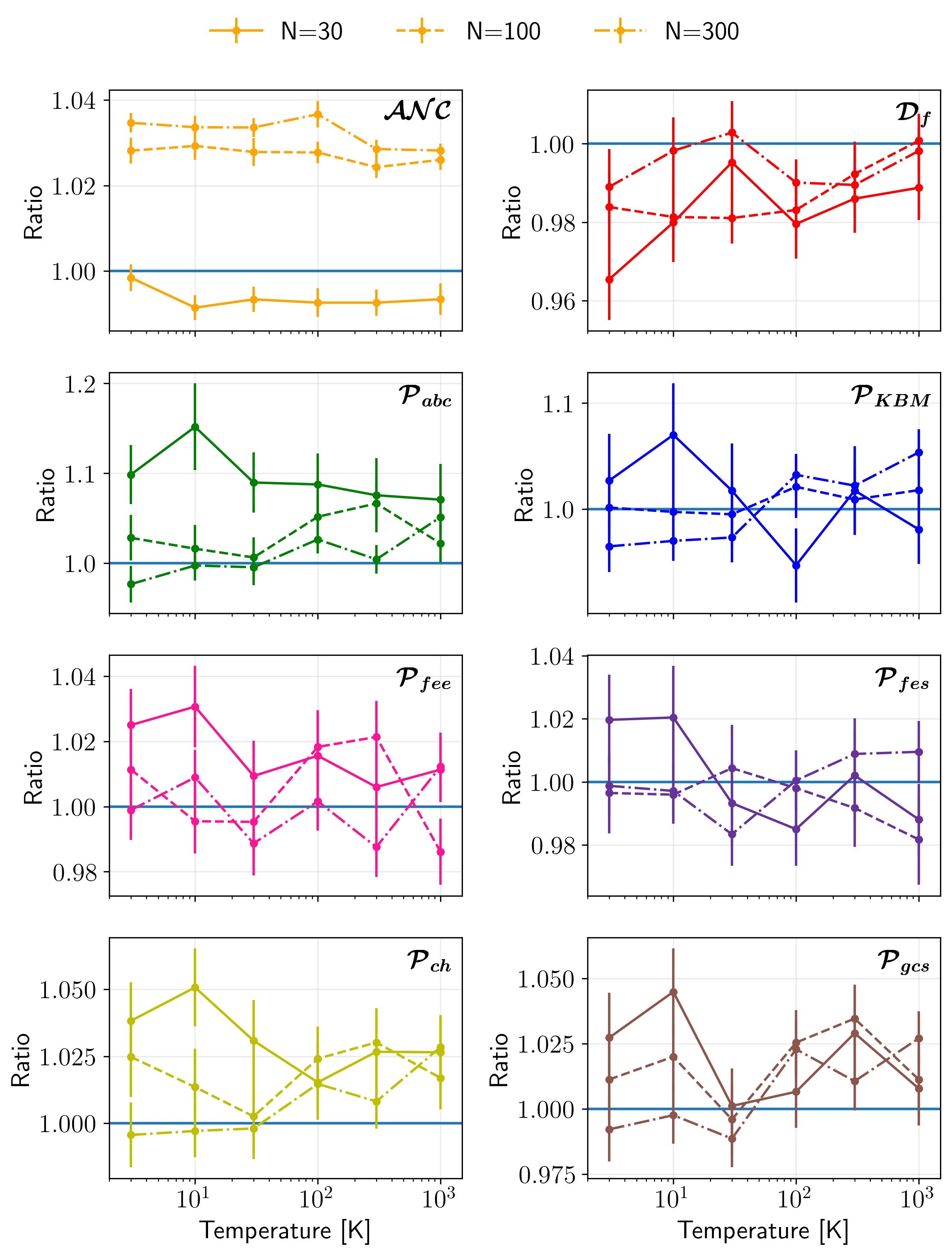}    

\caption{Ratio of constant to lognormal monomer radii distribution for each metric of aggregate structure over temperature for all aggregate sizes. Linestyles of all subfigures match the legend above panels (a) and (b). Solid line is $N=30$ monomers, dashed is $N=100$ monomers, and dashed-dot is $N=300$ monomers.} 
\label{fig:dist_comp}    
\end{figure*}

Additionally, DECCO now utilizes OpenMP, MPI, and GPU offloading to parallelize the computation of forces between monomer pairs. The results presented below utilize the parallelization capabilities, while the GPU offloading is still under testing.

\subsection{Aggregate Growth}
\label{subsec:aggGrowth}

Aggregates are grown one monomer at a time as in particle-cluster aggregation (PCA) and with sequential collisions as in \cite{Suyama2008}. We start with two monomers nearly touching as an initial target. Then, we choose a random direction to shoot a projectile monomer. The projectile is given a random offset in both dimensions of the plane perpendicular to the projectile direction. For both dimensions, the amount of offset is a uniformly random number between plus or minus the target aggregate's initial radius (again, calculated as the distance from the furthest monomer to the center of mass of the aggregate). A check is then performed to ensure the projectile monomer will collide with the target aggregate. If not, a new offset is chosen until a collision is confirmed for the initial direction and offset. Finally, the projectile is given a temperature-dependent initial velocity, and the simulation is carried out. The resulting, larger\footnote{While in principle it is possible that one or more monomers are ejected as a consequence of the collision, such a behavior was never observed for the monomer energies considered in the work.} aggregate becomes the target for the next simulation, and this process is repeated until the desired aggregate size is reached.

In the past, many Discrete Element Method (DEM) simulations have grown aggregates through sequential sticking. In sequential sticking, projectiles are placed on the target aggregate where an impact would have taken place. The difference between sequential sticking and sequential collisions is that in sequential collisions the collision of each projectile is fully simulated, and thus the incoming monomer may come to rest at a different location from the initial point of contact and/or intermediate restructuring can take place given sufficient energy in the collision.

Between collisions of projectile monomers and the target aggregate we zero out the velocity of the target's center of mass but carry over any angular momentum in the aggregate to the next simulation. In some cases, particularly for higher temperatures, the angular momentum of a fully grown aggregate is large enough that the aggregate would take a long time to fully relax. In order to speed up the relaxation we zero the angular momentum of the fully grown aggregate and continue the simulation before data processing. This does not appreciably change the structure of the aggregate, but rather allows for contacts to stay constant over time which is necessary for calculating the number of contacts reliably.

Given the size of our monomers, their velocities are assumed to be in thermal equilibrium with the diffuse gas around them\footnote{The assumption of thermal equilibrium is made to isolate and study the effects of gas temperature on aggregate structure. In some astrophysical environments magnetohydrodynamic turbulence can accelerate grains to higher speeds \cite{Yan2004}.}. In this case, motion is dominated by Brownian and Van der Waals dynamics \citep{Blum2004,Sarangi2018}. Thus, each projectile's velocity was chosen from a Maxwell-Boltzmann distribution for the given temperature and monomer mass. For the physics considered in this study, the temperature of the monomers/aggregates themselves are not important and do not affect sticking or restructuring. It is the temperature of the gas surrounding the monomers/grains which drives the physics. Example aggregates formed with DECCO for this study are shown in Figure~\ref{fig:agg_comp}.

Another common method of aggregate growth in the literature is cluster-cluster aggregation (CCA) \citep{Meakin1991}. In CCA, instead of adding single monomers, whole aggregates (sometimes copies of the same aggregate) are brought together to grow the aggregate. PCA and CCA are typically thought of as two extremes of aggregate growth. Current results show that PCA produces dense, more compact aggregates while CCA produces fluffy aggregates. Whether this assumption holds true when aggregates are grown with sequential collisions instead of sequential sticking requires a detailed study that is beyond the scope of this study. It is possible that larger projectiles in CCA would lead to more restructuring and a denser final aggregate compared to PCA, particularly for non-thermal impact velocities. CCA aggregates forming under conditions similar to the present study will be explored in a future work.

\subsection{Quantifying Aggregate Structure}
\label{subsec:structureQuantification}
Defining a quantity that can describe and differentiate possible structures a small, irregular aggregate can take is no easy task. Multiple definitions have been proposed, each with strengths and weaknesses. In this paper we test eight metrics to quantify the structure of an aggregate in order to thoroughly describe our results as well as to compare the metrics themselves. 

The first six metrics we discuss are metrics of porosity, or the free space in the aggregate. In general, the porosity is defined as $\mathcal{P} = 1-V_{eq}/V$, where $V_{eq}$ is the total volume of matter in the aggregate (known as the ``single equivalent sphere volume") and $V$ is the volume occupied by the aggregate. The single equivalent sphere volume is defined as:

\begin{equation}
    V_{eq}=\frac{4\pi}{3}\sum_i r_{i}^{3}
\end{equation}

\noindent where $r_i$ is the radius of the $i$th monomer in the aggregate and the sum is over all monomers in the aggregate. 

There are many ways to define $V$, the volume occupied by the aggregate. We discuss six distinct ways in the next six subsections. Figure~\ref{fig:porosityVisualization} visualizes how each porosity metric accounts for the volume occupied by an aggregate. Table~\ref{tab:toyPorosities} summarizes how these porosities behave when applied to idealized 3D shapes.

\begin{table}
\centering
\caption{Porosity values for idealized solid 3D geometries under each porosity definition considered in this work**.}
\label{tab:toyPorosities}

\begin{threeparttable}
\renewcommand{\arraystretch}{1.25}
\begin{tabular}{l c c c}
\toprule
Porosity Metric & Sphere & 3-axis Ellipsoid & Rectangular Prism \\
\midrule
$\mathcal{P}_{abc}$ & 0 & 0 & $0 < \mathcal{P}_{abc} < 1$ \\
$\mathcal{P}_{KBM}$ & 0 & $0 < \mathcal{P}_{KBM} < 1$ & $0 < \mathcal{P}_{KBM} < 1$ \\
$\mathcal{P}_{fee}$ & 0 & 0 & $0 < \mathcal{P}_{fee} < 1$ \\
$\mathcal{P}_{fes}$ & 0 & $0 < \mathcal{P}_{fes} < 1$ & $0 < \mathcal{P}_{fes} < 1$ \\
$\mathcal{P}_{ch}$  & 0\tnote{*} & 0\tnote{*} & 0 \\
$\mathcal{P}_{gcs}$ & 0 & $<1$ & $<1$ \\
\bottomrule
\end{tabular}

\begin{tablenotes}
\item[*] $\mathcal{P}_{ch}$ is strictly zero for a sphere and a three-axis ellipsoid only in the limit of infinitely many hull vertices. For any finite sampling, the resulting values are positive but arbitrarily close to zero.
\item[**] We do not provide proofs of these results, and rather use this table to help illustrate how these different porosities behave.

\end{tablenotes}
\end{threeparttable}
\end{table}

The last two metrics we discuss are the fractal dimension and average number of contacts. They are defined in the last two subsections of section~\ref{subsec:structureQuantification}.

\subsubsection{Equivalent Ellipsoid Porosity}
The first porosity definition relies on the volume of an ``equivalent ellipsoid" to define the volume occupied by the aggregate. This equivalent ellipsoid has the same mass and principal components of the moment of inertia tensor as the aggregate. To calculate this we follow \cite{Shen2008} and define the following dimensionless quantities:

\begin{equation}
    \alpha _i \equiv \frac{I_i}{0.4Mr^{2}_{eff}}
\end{equation}

\noindent where $\mathbf{I}$ is the moment of inertia tensor of the aggregate with eigenvalues $I_1 \ge I_2 \ge I_3$, $r_{eff}$ is the radius of the single equivalent sphere and is defined as $r_{eff} \equiv [3V_{eq}/({4\pi}) ]^{1/3}$ (also known as the effective radius), and $M \equiv \varrho _0 V_{eq}$. It is convenient to treat $\varrho_0$ as the monomer density. 

With this, and the requirement that the mass and principal components of $\mathbf{I}$ be equal for the aggregate and equivalent ellipsoid, the principle axes of the equivalent ellipsoid are

\begin{eqnarray}
    a &=& r_{eff}(\alpha _2 + \alpha _3 - \alpha _1)^{1/2} \\
    b &=& r_{eff}(\alpha _3 + \alpha _1 - \alpha _2)^{1/2} \\
    c &=& r_{eff}(\alpha _1 + \alpha _2 - \alpha _3)^{1/2}. 
\end{eqnarray}

\noindent If we define $R_{abc}=(abc)^{1/3}$, the porosity based on the equivalent ellipsoid is

\begin{equation}
    \mathcal{P}_{abc} \equiv 1-\left(\frac{r_{eff}}{R_{abc}}\right)^{3}.
\end{equation}

For a more in-depth derivation of $R_{abc}$ see \citet{Shen2008}. Panel (c) of Figures~\ref{fig:const_measures} and~\ref{fig:lognorm_measures} shows the value of $\mathcal{P}_{abc}$ vs temperature for constant and lognormal monomer size distributions, respectively. 

Note that the volume equivalent ellipsoid does not necessarily enclose the whole aggregate. Panel (a) in Figure~\ref{fig:porosityVisualization} shows an example of the volume equivalent ellipsoid for one of our aggregates. 

\subsubsection{Gyration Radius Based Porosity}

The second definition of volume occupied by an aggregate is calculated with a radius proportional to the radius of gyration, $R_{KBM}$. As described again by \cite{Shen2008}, for an aggregate with uniform density, $R_{KBM}=(5/3)^{1/2}R_{gyr}$. The radius of gyration for an aggregate is defined as:

\begin{equation}
    R_{gyr} = \sqrt{\frac{\sum_i m_i ||\vec{x}_{i} - \vec{x}_{COM}||^2}{M_{tot}}}    
\end{equation}

\noindent where $M_{tot}$ is the total mass of the aggregate, $m_i$ is the mass of the $i$th monomer, $\vec{x}_i$ is the position of the center of the $i$th monomer, and $\vec{x}_{COM}$ is the position of the center of mass of the aggregate. In our case, we use \cite{Shen2008}'s equivalent definition of $R_{KBM}$:

\begin{equation}
    R_{KBM} \equiv \left(\frac{\alpha _1 + \alpha _2 + \alpha _3}{3}\right)^{1/2} r_{eff}.
\end{equation}

\FloatBarrier

\noindent Thus, the second porosity follows \citet{Kozasa1992}:
\begin{equation}
    \mathcal{P}_{KBM} \equiv 1-\left(\frac{r_{eff}}{R_{KBM}}\right)^{3}.
\end{equation}

\FloatBarrier

\noindent Note that $\mathcal{P}_{KBM}$ and the bulk density $\rho/\rho_0$, from \cite{Suyama2008} and \cite{Tanaka2023} are directly related with the expression $\mathcal{P}_{KBM} = 1-\rho/\rho_0$.
Panel (d) of Figures~\ref{fig:const_measures} and~\ref{fig:lognorm_measures} shows the value of $\mathcal{P}_{KBM}$ vs temperature for constant and lognormal monomer size distributions, respectively.

Like for $\mathcal{P}_{abc}$, $\mathcal{P}_{KBM}$ does not necessarily enclose the whole aggregate, shown in panel (b) of Figure~\ref{fig:porosityVisualization}.

\subsubsection{Fully Enclosing Sphere Porosity}

Probably the simplest definition of the volume occupied by an aggregate is a sphere that fully encloses the aggregate. Enclosing spheres have been used to describe various morphologies of dust for use in Discrete Dipole Approximation calculations \citep{Min2006,Zubko2015,Botet2025}. In this study we use the minimum volume enclosing sphere. 

To compute the enclosing sphere, we solve the optimization problem

\begin{equation}
\min_{c,\,R_{fes}}\; R_{fes} 
\quad \text{s.t.} \quad
\|c - x_i\|_2 + r_i \le R_{fes},\; \forall i 
\end{equation}

\noindent where $c$ is the center of the fully enclosing sphere, $R_{fes}$ is its radius, $\|\cdot\|_2$ denotes the Euclidean norm, and $i$ is over all monomers in the aggregate.

\noindent Using this definition, the porosity becomes

\begin{equation}
    \mathcal{P}_{fes} \equiv 1-\left(\frac{r_{eff}}{R_{fes}}\right)^{3}.
\end{equation}

This method of defining aggregate volume is only applicable to aggregates that are mostly spherical. If aggregates are not spherical, an enclosing sphere can vastly overestimate aggregate volume.

Panel (f) of Figures~\ref{fig:const_measures} and~\ref{fig:lognorm_measures} shows the value of $\mathcal{P}_{fes}$ vs temperature for constant and lognormal monomer size distributions, respectively.

\subsubsection{Fully Enclosing Ellipsoid Porosity}
A natural progression from a fully enclosing sphere is a fully enclosing ellipsoid. Calculating a fully enclosed ellipsoid is a well studied problem in computer graphics. In this case, it is often referred to as the Minimum Volume Enclosing Ellipsoid (MVEE) or Löwner-John ellipsoid.

This problem is often written as \citep{Todd2007,Bowman2023}

\begin{equation}
\label{eq:MVEE}
\begin{aligned}
\min_{c,H} \quad & -\log\det(H) \\
\text{s.t.} \quad 
& (x_j - c)^{\top} H (x_j - c) \le 1,\qquad \forall j \\[2mm]
& H \succ 0.
\end{aligned}
\end{equation}

\noindent where $H$ is a symmetric positive definite matrix containing information about the ellipsoid's principal axes and rotation matrix, $\log\det(H)$ is the natural log of the determinant of $H$, $c$ is now the center of the ellipsoid, and $j$ is over all points to be enclosed by the ellipsoid. Unlike for the fully enclosing sphere, the fully enclosing ellipsoid is not trivial to extend to enclosing spheres with a finite radius. Thus, we use Fibonacci sampling to place 8192 points on the surface of each monomer to ensure adequate coverage. To solve this problem we use the \texttt{mvee2} function from the code written for \cite{Bowman2023}.

Thus, given the principle axes of the fully enclosing ellipsoid are $p \ge q \ge s$ and $R_{fee}=(pqs)^{1/3}$, the porosity based on the fully enclosing ellipsoid is

\begin{equation}
    \mathcal{P}_{fee} \equiv 1-\left(\frac{r_{eff}}{R_{fee}}\right)^{3}.
\end{equation}

We are not aware of any study which uses this definition of porosity, but it is useful as a comparison to the fully enclosing sphere and other porosity metrics in this study.

Panel (e) of Figures~\ref{fig:const_measures} and~\ref{fig:lognorm_measures} shows the value of $\mathcal{P}_{fee}$ vs temperature for constant and lognormal monomer size distributions, respectively.

\subsubsection{Convex Hull Porosity}
Since aggregates are themselves irregular in shape, a natural choice to describe their volume is a hull. If one wishes to use a hull that allows for concavity, the problem becomes choosing what sized cavity to include/exclude from the hull. Since this choice will change the hull, it is preferable to not make such a choice if possible. If an aggregate is compact enough, a convex hull is a good approximation for its shape. Convex hulls have been used to define aggregate volume in both experimental \citep{Kohout2014} and numerical \citep{Gunkelmann2016,Bandyopadhyay2023} studies of cosmic dust.

To calculate our convex hulls we use Fibonacci sampling to place 64 points on each monomer's surface, and pass those points to the \texttt{scipy.spatial.ConvexHull} function from the SciPy library \citep{Virtanen2020}, which is built on the Qhull algorithm \citep{Barber1996}. We chose 64 points per monomer in this case as there was almost no difference between the hull volumes calculated from 64 and 8192 points. This function gives the volume of the convex hull ($V_{ch}$). Keeping with our previous notation, we can use this volume to define a radius $R_{ch}=\sqrt[3]{\frac{3}{4\pi}V_{ch}}$, and the porosity becomes

\begin{equation}
    \mathcal{P}_{ch} \equiv 1-\left(\frac{r_{eff}}{R_{ch}}\right)^{3}.
\end{equation}

Panel (g) of Figures~\ref{fig:const_measures} and~\ref{fig:lognorm_measures} shows the value of $\mathcal{P}_{ch}$ vs temperature for constant and lognormal monomer size distributions, respectively.

\subsubsection{Geometric Cross Section Porosity}
An aggregate's size can also be described using its geometric cross section. The geometric cross section can be thought of as the area of a shadow projected onto a plane by the aggregate when illuminated with a beam of parallel rays normal to the plane.

We follow the method in \cite{Suyama2012} to calculate the geometric cross section ($S$). In this method, a grid is defined on the plane in which to calculate the geometric cross section. The grid size is $0.0055r_{min}$, where $r_{min}$ is the radius of the smallest monomer in the aggregate. The area of grid spaces the shadow falls on are then summed to obtain the geometric cross section. This value $S$ is then averaged over 30 random directions.

The geometric cross section can be used to find the radius of a circle which has the same surface area, $R_{gcs}=\sqrt{S/\pi}$. This radius is sometimes known as the ``projected surface equivalent radius" of an aggregate \citep{Ormel2009}. A porosity based on this radius is then

\begin{equation}
    \mathcal{P}_{gcs} \equiv 1-\left(\frac{r_{eff}}{R_{gcs}}\right)^{3}.
\end{equation}

We also include a table of the value of $S$ for each aggregate size, temperature, and monomer size distribution in the appendix.

Panel (h) of Figures~\ref{fig:const_measures} and~\ref{fig:lognorm_measures} shows the value of $\mathcal{P}_{gcs}$ vs temperature for constant and lognormal monomer size distributions, respectively.

\subsubsection{Fractal Dimension}
Another metric of aggregate structure is the fractal dimension. For the fractal dimension we used the box counting dimension, or, in three dimensions, the cube counting dimension. The cube counting dimension places an object (in our case, an aggregate) in a 3D grid and then counts the number of grid spaces that contain any part of the aggregate. Repeat this for different grid sizes and one can determine how the number of grid spaces containing the aggregate changes with grid size. Formally, the fractal dimension $D_f$ is defined as:

\begin{equation}
    D_f \equiv \lim_{l \to 0} \frac{log(\mathcal{N}(l))}{log(1/l)}
\end{equation}

\noindent where $\mathcal{N}(l)$ is the number of grid spaces of size $l$ containing any portion of the object. In practice, our aggregates are not true fractals and are only fractal-like at a certain scale. Thus, we calculate the fractal dimension by plotting $log(\mathcal{N}(l))$ vs $log(1/l)$ for several box sizes, starting at the size of the aggregate and halving until the box size is less than or equal to the radius of the smallest monomer. We then fit a line to the four points on this plot that give the most linear fit\footnote{This is done because aggregates are only fractal over a range of box sizes. By taking the four most linear points we leave out box sizes outside of this range that would bias our values of $D_f$.}. The fractal dimension reported is the slope of this most linear fit. Panel (b) of Figures~\ref{fig:const_measures} and~\ref{fig:lognorm_measures} shows the value of fractal dimension vs temperature for constant and lognormal monomer size distributions, respectively.

\subsubsection{Average Number of Contacts}
The average number of contacts ($\mathcal{ANC}$) is simply the average number of other monomers each monomer is touching. If $r_i$ and $r_j$ are the radii of the $i$th and $j$th monomers, they are in contact if the distance between the position of their centers $\vec{x}_i$ and $\vec{x}_j$ is less than or equal to the sum of their radii. Panel (a) of Figures~\ref{fig:const_measures} and~\ref{fig:lognorm_measures} shows the value of average number of contacts vs temperature for constant and lognormal monomer size distributions, respectively.

\section{Results}
\label{sec:results}

We generated aggregates containing $N=30$, $100$, and $300$ monomers at temperatures of $3$, $10$, $30$, $100$, $300$, and $1000$~K. For each combination of size and temperature, we considered both constant and lognormal monomer size distributions. Every simulation was repeated 30 times using different random seeds. Representative aggregates for all sizes, temperatures, and monomer size distributions are shown in Figure~\ref{fig:agg_comp}.

For each simulated aggregate we computed the eight structural metrics. The final results were obtained by averaging each metric across the 30 simulations for each parameter set, reducing the influence of random noise.

\subsection{Structural response to aggregate size}

We find that there is a statistically significant difference between structural parameter values for different aggregate sizes for all metrics tested. This is shown in Figures~\ref{fig:const_measures} and \ref{fig:lognorm_measures}, with the single exception of the fractal dimension for lognormal aggregates in panel (b) of Figure~\ref{fig:lognorm_measures}, where the average values for $N=100$ overlap and cross over the $N=30$ and $300$ monomer aggregate average values.

For all metrics, a larger aggregate has a larger value of all structural parameters, irrespective of temperature. Given that, intuitively, a high porosity indicates a less dense aggregate, and a high average number of contacts/fractal dimension indicates a more dense aggregate, this is not a straightforward result. It highlights differences in how these metrics quantify structure. 

One important difference among the metrics used is the extent to which they rely on local versus global properties of the aggregate. The average number of contacts, for example, is based entirely on local properties that are then averaged over the whole aggregate. A single monomer can only feel the monomers directly around it. While monomers in the bulk are surrounded by other potential contacts, those on the surface can only contact other monomers on one side. Thus, as aggregates get larger, surface effects on this metric are weighted less, which we believe explains the observed trend. 

The porosity metrics on the other hand, compare two volumes, one which describes the amount of solid material and the other which describes the space an aggregate occupies. As such, porosities are intrinsically defined for a finite aggregate and are constructed to be a global measure. The increasing porosity with size that we observe
may be related to the fact that a minimum number of monomers are required to create voids inside the aggregate. It is expected that porosity will stabilize to a constant value beyond a certain size. Such a size, however, was beyond our ability to simulate in a reasonable computational time with the current version of DECCO on available hardware.

Fractal dimension sits in between, relying on a multi-scale analysis of the aggregate and therefore capturing both the local and global morphologies. It is not unexpected, therefore, that it appears to be the least sensitive to size (see panel (b) in Figures~\ref{fig:const_measures} and~\ref{fig:lognorm_measures}). Fractal dimension is, however, the most computationally intensive parameter to assess. In addition, a minimum number of monomers is required for the aggregate to cover at least a factor of 8 in size\footnote{Since our method for measuring the fractal dimension involves a fit across 4 box sizes (see Section~\ref{subsec:structureQuantification}), a robust measurement requires the aggregate to be at least $2^3=8$ times larger than the radius of the smallest monomer involved.} between the smallest monomer and the whole aggregate. For this reason, the fractal dimension of our smallest aggregates are calculated from the minimum of four box sizes, and should be considered estimates rather than true measures.

\subsection{Structural Response to Temperature}
\label{subsec:responseToTemp}

Each structural metric we tested vs.\@ temperature is shown in Figure~\ref{fig:const_measures} for a constant monomer size distribution and Figure~\ref{fig:lognorm_measures} for lognormal. Our results show a statistically significant increase in aggregate density with increasing temperature in nearly all cases of aggregate size and monomer size distribution. This is shown in Figure~\ref{fig:method_comp}, where we fit the data in Figure~\ref{fig:const_measures} and \ref{fig:lognorm_measures} to straight lines and report the absolute value of their slopes, which we call the sensitivity to temperature\footnote{Note that for the number of contact and fractal dimension cases the slopes are positive, while for the porosities the slopes are negative.}.

In general the sensitivity to temperature is greater for larger aggregates. The exception to this is $\mathcal{P}_{KBM}$ and the fractal dimension for constant aggregates (Figure~\ref{fig:const_measures}), as well as the fractal dimension for lognormal aggregates (Figure~\ref{fig:lognorm_measures}), where the sensitivity to temperature is consistent over aggregate size.

\subsection{Comparison between monomer size distributions}
\label{subsec:CompareSizeDist}

Figure~\ref{fig:dist_comp} shows the ratio of the constant distribution metric values (from Figure~\ref{fig:const_measures}) to the lognormal distribution metric values (from Figure~\ref{fig:lognorm_measures}). It is to be expected that a non-constant monomer size distribution should produce more dense aggregates compared to a constant distribution, since the smaller monomers can fill in between gaps left by the larger ones. Our results in Figure~\ref{fig:dist_comp} support this expectation for the fractal dimension where the ratio is below one, and porosities $\mathcal{P}_{abc}$, $\mathcal{P}_{ch}$, and $\mathcal{P}_{gcs}$ where the ratio is above one. In the cases of $\mathcal{P}_{KBM}$, $\mathcal{P}_{fee}$, and $\mathcal{P}_{fes}$ instead, the ratio is centered around one,  indicating that these metrics are not sensitive to differences in the monomer size distribution. 

For the average number of contacts this result is less clear. The ratio of the average number of contacts for $N=100$ and $300$ monomers is above unity, indicating the constant size distribution aggregates are more dense. In contrast, the ratio for $N=30$ monomers is below unity, indicating the constant size distribution aggregates are less dense at small sizes. We believe this is another shortcoming in the average number of contacts metric, which is further discussed in Section~\ref{sec:summary}. 

When looking at temperature sensitivity across monomer size distributions, we find that lognormal aggregates have a structure that depends more on temperature than constant aggregates. This is particularly evident for the larger aggregates ($N=300$ monomers) in Figure~\ref{fig:method_comp}, in which the orange symbols in the bottom panel are consistently higher than those in the top panel. On the other hand, small ($N=30,100$ monomers) lognormal aggregates respond similarly to temperature than constant aggregates. This might be related to the fact that a minimum number of monomers is needed to sample the lognormal distribution widely enough to produce a significant difference with constant sized aggregates.

\begin{figure}
\includegraphics[width=0.99\linewidth,trim={140 35 137 40},clip]{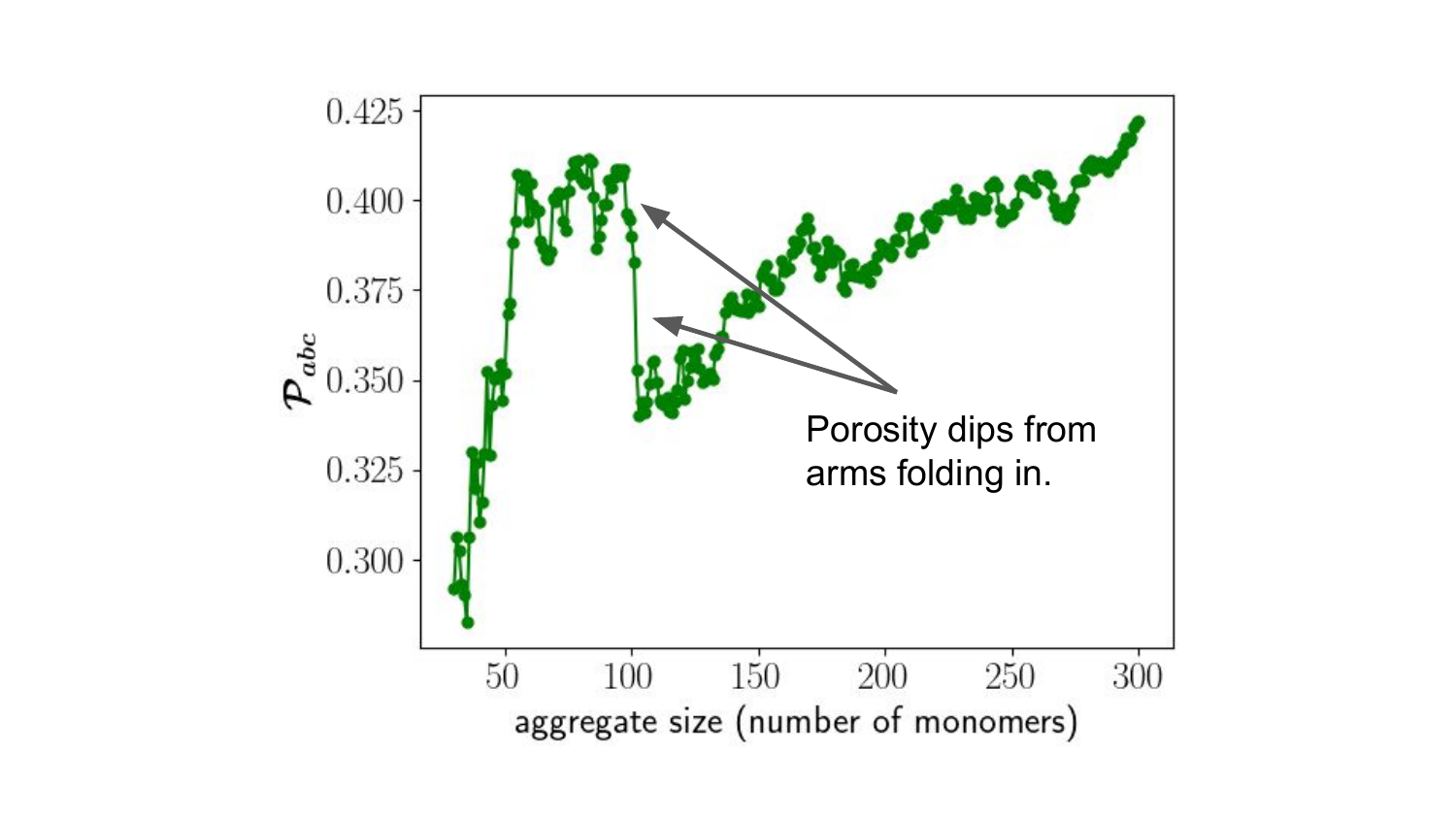}
    \caption{$\mathcal{P}_{abc}$ vs aggregate size for a single growing lognormal aggregate at 3K. 
    \label{fig:porosityOverTime}}
\end{figure}

\subsection{Structural Evolution Over Time}\label{subsec:responseOverTime}

We also followed the porosity evolution of $N=300$ monomer aggregates as they grew to investigate whether the porosity evolves smoothly or if a few high energy impacts cause sudden drops in porosity. In general we found that the porosity evolves sharply and chaotically before 100 monomers and mostly smoothly after 100 monomers for both high and low temperature aggregates. The main difference between the evolution of high-temperature and low-temperature aggregates is that some of the low temperature aggregates experience sudden drops in porosity, while this feature is not seen for high temperature aggregates. Figure~\ref{fig:porosityOverTime} shows $\mathcal{P}_{abc}$ vs aggregate size for a single lognormal aggregate formed at 3~K. The two indicated dips in $\mathcal{P}_{abc}$ correspond to arms\footnote{An ``arm" is a collection of monomers that juts out from the center of an aggregate.} folding into the aggregate due to a projectile monomer. This feature is not seen for high temperature aggregates that do not form prominent arms.

\begin{table}[!t]
    \centering
    \begin{tabular}{lcc}
        \toprule
        & Sequential Sticking & Sequential Collision at 3 K \\
        \midrule
        \multicolumn{3}{l}{\textbf{Constant monomer size distribution}} \\
        $\mathcal{P}_{abc}$ & $0.685 \pm 0.007$ & $0.470 \pm 0.006$ \\
        $\mathcal{P}_{KBM}$ & $0.776 \pm 0.007$ & $0.527 \pm 0.009$ \\
        $\mathcal{P}_{fee}$ & $0.862 \pm 0.004$ & $0.763 \pm 0.005$ \\
        $\mathcal{P}_{fes}$ & $0.922 \pm 0.003$ & $0.823 \pm 0.005$ \\
        $\mathcal{P}_{ch}$  & $0.724 \pm 0.005$ & $0.576 \pm 0.005$ \\
        $\mathcal{P}_{gcs}$ & $0.669 \pm 0.005$ & $0.531 \pm 0.004$ \\
        \midrule
        \multicolumn{3}{l}{\textbf{Lognormal monomer size distribution}} \\
        $\mathcal{P}_{abc}$ & $0.686 \pm 0.005$ & $0.481 \pm 0.007$ \\
        $\mathcal{P}_{KBM}$ & $0.760 \pm 0.006$ & $0.547 \pm 0.010$ \\
        $\mathcal{P}_{fee}$ & $0.860 \pm 0.003$ & $0.764 \pm 0.005$ \\
        $\mathcal{P}_{fes}$ & $0.915 \pm 0.003$ & $0.824 \pm 0.006$ \\
        $\mathcal{P}_{ch}$  & $0.727 \pm 0.004$ & $0.578 \pm 0.005$ \\
        $\mathcal{P}_{gcs}$ & $0.663 \pm 0.003$ & $0.536 \pm 0.005$ \\
        \bottomrule
    \end{tabular}
    \caption{Average of 30 aggregates ($N=300$ monomers) for different porosity definitions under sequential sticking and sequential collisions at 3 K, shown separately for constant and lognormal monomer size distributions.}
    \label{tab:seqStickSeqColl}
\end{table}

\subsection{Comparison to Sequential Sticking}\label{subsec:sequentialSticking}

In addition to growing aggregates by sequential collisions we grew 30, $N=300$ monomer aggregates with sequential sticking for both a constant and lognormal monomer size distribution. These aggregates grow by sticking (rather than colliding) single monomers on the target from a random direction. These additional simulations allowed us to study how these new aggregates compare to the ones we grew with DECCO. In order for a fair comparison, these aggregates are also allowed time to relax into a final state like has been done for the sequential collision aggregates.

Since sequential sticking can loosely be thought of as sequential collisions at a temperature of absolute zero (depositing zero energy due to the ``collision"), it is no surprise that these aggregates are very porous. Porosity values for the sequential sticking aggregates and 3K sequential collision aggregates are shown in Table~\ref{tab:seqStickSeqColl}.

\section{Summary and Discussion} \label{sec:summary}

We have used our discrete element code, DECCO \citep{Guidos2025}, to study the structure of dust aggregates formed by coagulation in an environment consistent with a young SNR. After forming dust aggregates with the PCA methodology, we used eight different metrics to quantify their structure. These are six porosities ($\mathcal{P}_{abc}$, $\mathcal{P}_{KBM}$, $\mathcal{P}_{fee}$, $\mathcal{P}_{fes}$, $\mathcal{P}_{ch}$, and $\mathcal{P}_{gcs}$), the average number of contacts, and the fractal dimension. 

We find that aggregates tend to form a denser structure with increasing temperature and when monomers of different sizes are involved. We also find the structure of the largest aggregates depends more on temperature, particularly for a non-uniform monomer size distribution. These results are relevant to the study of dust resilience at the crossing of the reverse shock \citep{Bianchi2007,Nozawa2007,Slavin2020,Kirchschlager2024}. 

So far, however, reverse shock dust destruction studies have predominantly used single monomers, and not aggregates, in their simulations \citep{Nozawa2007,Bianchi2007,Slavin2015,Bocchio2016,Micelotta2016,Micelotta2018,Kirchschlager2019,Slavin2020,Kirchschlager2023}. If aggregates are treated, dust destruction may look different. It is difficult to predict how this difference would play out. If we consider large grains as aggregates instead of large monomers, previous studies could be underestimating dust destruction. If instead we compare aggregates formed at high vs low temperatures the results are equally unpredictable. On one hand, dense aggregates formed at a higher temperature and likely made of a distribution of monomer sizes (see, e.g. Figures~8 and~9 in \citealt{Fallest2011}), could be seen as more resilient to disruption than dust formed in colder environments. On the other hand, lighter/fluffier aggregates would couple more effectively with the gas dynamics and, as a consequence, be exposed to sputtering for a shorter amount of time. In this case, composite grains with larger cross sections would suffer less sputtering, and be more resilient than the large monomers case. Non-linear effects may also play a role. Aggregate particles respond differently to grain-grain collisions than large single monomers, and can be disaggregated into a collection of free small monomers through mechanical and radiative torques at much lower energies then their monolithic counterparts. It is therefore important to properly fold the aggregate properties discussed here in the hydrodynamic models of SNR evolution like those cited above. Likewise, it is critical to understand whether the aggregation process proceeds from coagulation to coalescence or not by the time of the reverse shock.

We have also used our results to assess the robustness and usefulness of different density/compactness metrics for aggregates. Despite its computational simplicity, we found the average number of contacts to be the least reliable measure. It is strongly affected by surface effects, which bias it with aggregate size, and it behaves counterintuitively when monomers of different sizes are involved. All the other metrics yielded coherent values, the clearest output coming from $\mathcal{P}_{abc}$, $\mathcal{P}_{ch}$, $\mathcal{P}_{gcs}$, and the fractal dimension as these metrics follow expectations for differentiating aggregates with constant and polydisperse monomers (cfr. Figure~\ref{fig:dist_comp}).

Despite the promising results, our study has some limitations and additional research is needed to better understand dust aggregate structure and their fate in the thermodynamical and dynamical conditions of a young SNR. First, our study only considers spherical monomers, which might be a grossly oversimplified representation of a true dust grain shape. Second, we adopt a fairly simple model for cohesion forces that is based on the Van der Waals potential. A Tersoff potential, implemented, e.g., in the LAMMPS suite \citep{LAMMPS}, might be a better representation for covalent bonds. Alternatively, more advanced models focus on the contact surface and include twisting and activation forces (the JKR model, used, e.g., in \citealt{Dominik1997,Wada2007}). These models, however, are strongly dependent on the assumed monomer shape, and may make the choice of spherical monomers even more fraught. Additional complications may be brought about by the transition from coagulation to coalescence, with the formation of a unique grain from the aggregation of smaller units. Before any firm conclusion on SN dust resilience can be drawn, all these effects and techniques need to be studied in detail and better understood. 

\begin{acknowledgements}
We would like to thank the anonymous reviewer's thoughtful comments which contributed greatly to improving the paper. This project was supported by NASA APRA award 80NSSC19K0330 and in part through NASA and Oregon Space Grant Consortium, cooperative agreement 80NSSC20M0035. LK also acknowledges financial support from the ARCS Foundation Oregon, as well as access to facilities and help developing the DECCO code from Dr. Doru Thom Popovici and Dr. Mauro Del Ben with the Applied Computing for Scientific Discovery group at LBNL. 
\end{acknowledgements}

\bibliography{main}

\appendix

\section{Comparison between porosity metrics}

\begin{figure}[h]
    \centering
    \includegraphics[width=\textwidth]{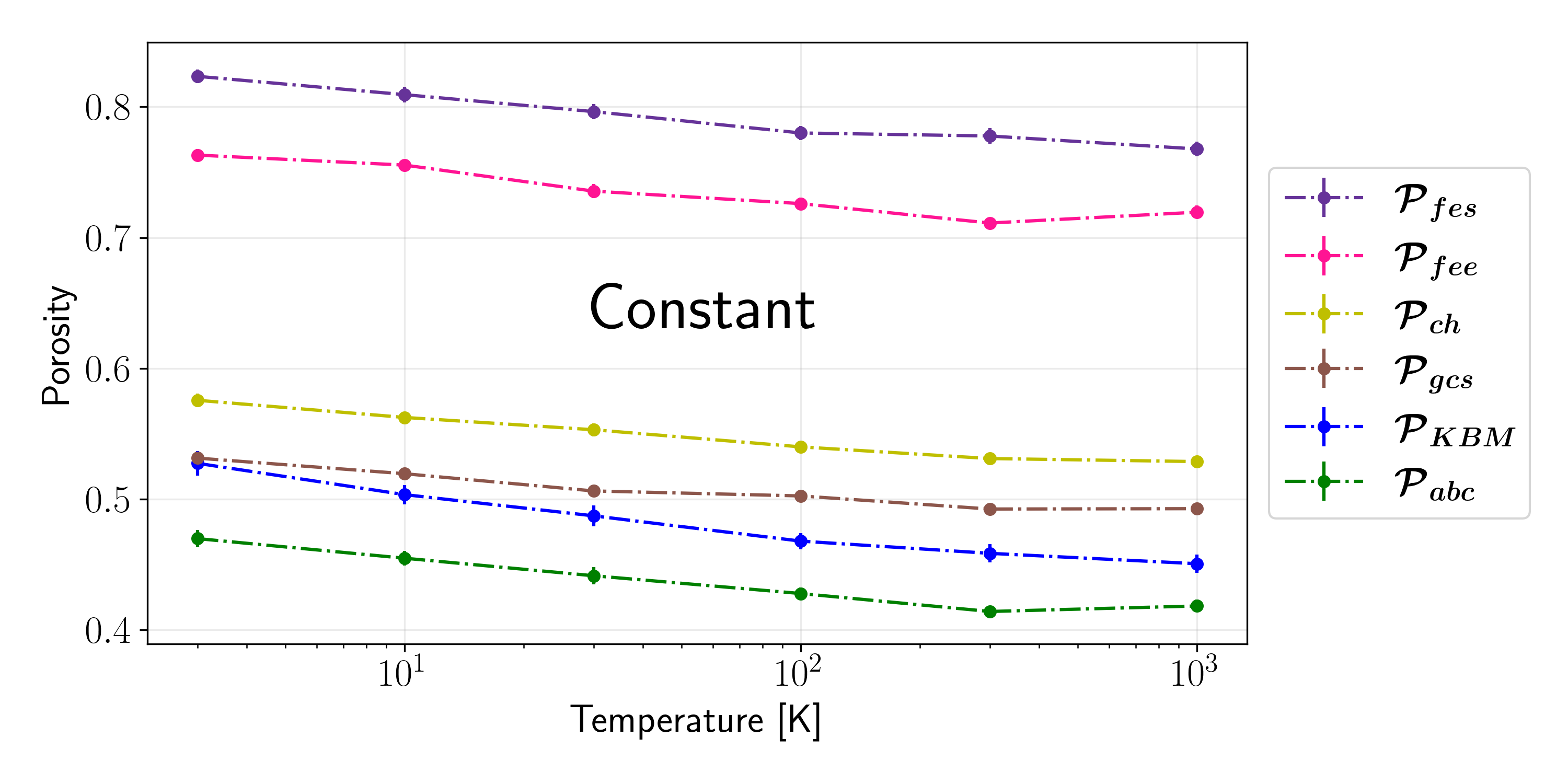}
    \caption{Comparison of each porosity metric tested vs temperature for $N=300$, constant monomer size distribution aggregates. The data shown is the same as the $N=300$ monomer porosity data in Figure~\ref{fig:const_measures}, but on the same axes.}
    \label{fig:constPorosities}
\end{figure}

\begin{figure}[h]
    \centering
    \includegraphics[width=\textwidth]{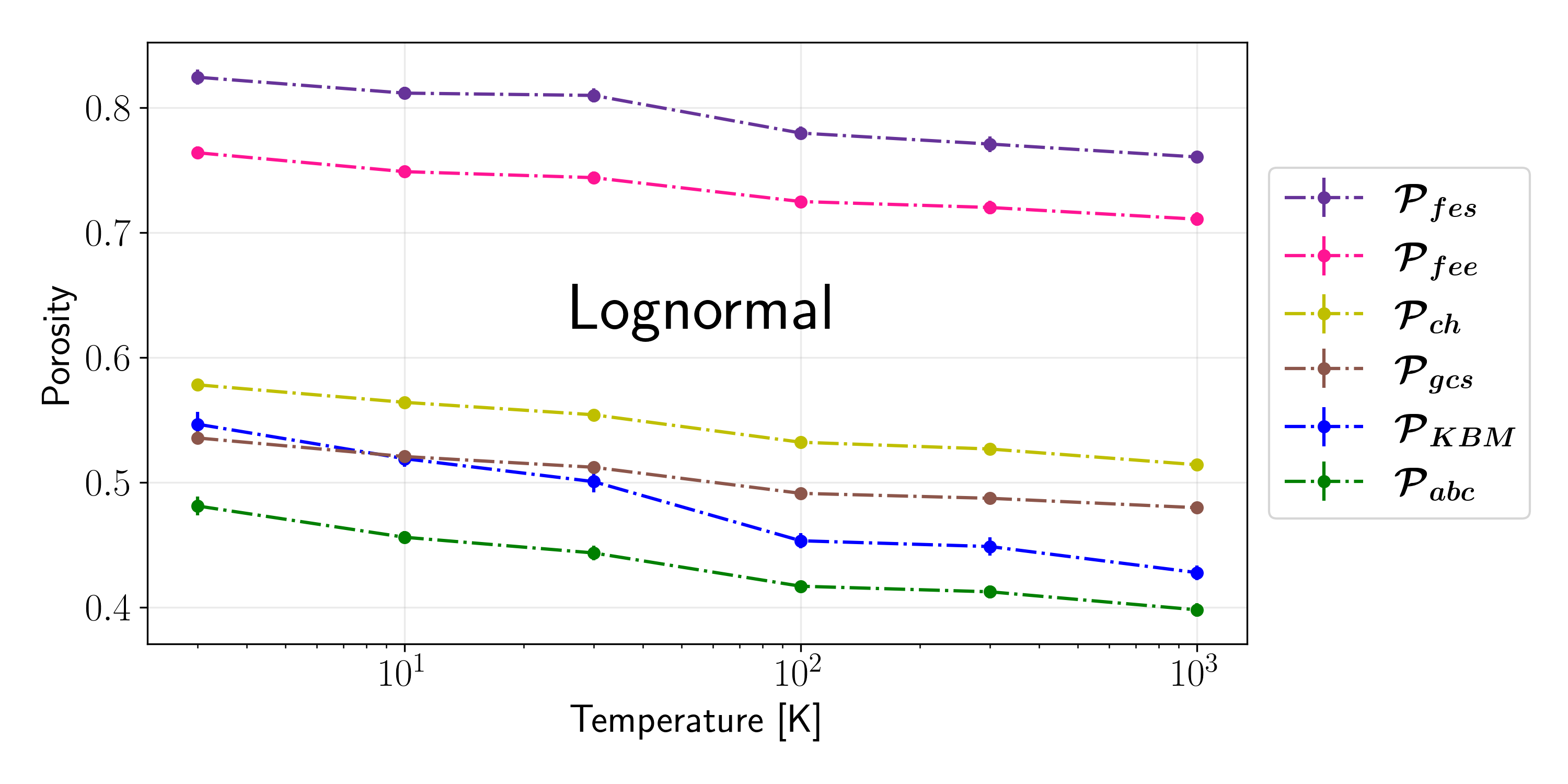}
    \caption{Comparison of each porosity metric tested vs temperature for $N=300$, lognormal monomer size distribution aggregates. The data shown is the same as the $N=300$ monomer porosity data in Figure~\ref{fig:lognorm_measures}, but on the same axes.}
    \label{fig:lognormPorosities}
\end{figure}

As discussed in Section~\ref{subsec:structureQuantification}, porosity metrics differ primarily in how they define the total volume occupied by an aggregate. The metrics explored in this study can be grouped into two categories: interaction-size metrics ($\mathcal{P}_{fes}$, $\mathcal{P}_{fee}$, $\mathcal{P}_{ch}$, and $\mathcal{P}_{gcs}$) and internal-structure metrics ($\mathcal{P}_{abc}$ and $\mathcal{P}_{KBM}$). Each category has distinct strengths and weaknesses, depending on the physical context in which porosity is being measured.

The interaction size of an aggregate is important for grain-grain collisions and interactions with electromagnetic radiation, but the choice of metric should depend on how dense or fractal the aggregate structure is.

The most intuitive interaction-size–based porosities are those defined by the fully enclosing sphere and ellipsoid. By construction, these metrics never underestimate the size of an aggregate and therefore provide an upper bound on its potential interactions. Their primary limitation, however, is that they can substantially overestimate the aggregate’s effective size. Figures~\ref{fig:constPorosities} and~\ref{fig:lognormPorosities} show the porosity as a function of temperature for $N=300$ monomer aggregates with constant and lognormal monomer size distributions, respectively. In both cases, the fully enclosing sphere and ellipsoid porosities are significantly larger than the other porosity measures.

For the aggregates studied here, the convex hull provides a more accurate estimate of interaction size than the fully enclosing sphere or ellipsoid, as the aggregates are relatively compact and lack significant concave features. In contrast, for highly fractal aggregates, a convex hull may enclose substantial void regions. Whether such voids should be considered part of the effective interaction size depends on both their spatial scale and the physical nature of the interaction.

The geometric cross section excludes more void space from the interaction ``size" than the convex hull. This can be appropriate when the interaction involves particles or radiation with characteristic scales much smaller than the voids themselves. However, it becomes less appropriate when the interacting object is comparable in size to, or larger than, these voids, in which case the voids should reasonably be considered part of the aggregate’s effective size.

For the dense aggregates considered here, the convex hull and geometric cross section porosities behave quantitatively very similarly in Figures~\ref{fig:constPorosities} and~\ref{fig:lognormPorosities}, and are also comparable to $\mathcal{P}_{abc}$ and $\mathcal{P}_{KBM}$.

When the primary interest is the internal structure of an aggregate rather than its interaction size, $\mathcal{P}_{abc}$ and $\mathcal{P}_{KBM}$ provide more informative metrics. This is because a wide range of distinct internal configurations can occupy the same enclosing sphere, enclosing ellipsoid, convex hull, or geometric cross section, whereas far fewer configurations satisfy the geometric constraints inherent to $\mathcal{P}_{abc}$ and $\mathcal{P}_{KBM}$.

As an extreme example, consider an enclosing sphere. Monomers could be distributed in a single straight line within the sphere or evenly throughout its volume. In both cases, the same enclosing sphere describes the aggregate, but the equivalent ellipsoid and gyration radius would be vastly different. On the other hand, as visualized in Figure~\ref{fig:porosityVisualization}, $\mathcal{P}_{abc}$ and $\mathcal{P}_{KBM}$ do not necessarily enclose the aggregate and therefore lose information about how the aggregate will interact. Thus, it is debatable how well these metrics capture the volume occupied by non-spherical aggregates.

\section{tabulated geometric cross sections}

\begin{table}[htbp]
\centering
\caption{Average geometric cross section as a function of temperature.
Values are shown as mean $\pm$ standard deviation over 30 realizations in units of $\mu\mathrm{m}^2$.}
\label{tab:gcs_vs_temp}

\textbf{Constant aggregates}

\vspace{0.5ex}

\begin{filecontents*}{gcsTable-constrelax.csv}
N,T3K,T10K,T30K,T100K,T300K,T1000K
30,0.444 ± 0.012,0.442 ± 0.012,0.438 ± 0.013,0.433 ± 0.010,0.440 ± 0.013,0.432 ± 0.012
100,1.053 ± 0.042,1.059 ± 0.033,1.036 ± 0.035,1.041 ± 0.033,1.029 ± 0.031,1.017 ± 0.025
300,2.337 ± 0.075,2.297 ± 0.066,2.257 ± 0.074,2.244 ± 0.060,2.214 ± 0.054,2.215 ± 0.054
\end{filecontents*}
\pgfplotstabletypeset[
  col sep=comma,
  string type,
  columns={N,T3K,T10K,T30K,T100K,T300K,T1000K},
  columns/N/.style={column name={$N$ (monomers)}},
  columns/T3K/.style={column name={$3\,\mathrm{K}$}},
  columns/T10K/.style={column name={$10\,\mathrm{K}$}},
  columns/T30K/.style={column name={$30\,\mathrm{K}$}},
  columns/T100K/.style={column name={$100\,\mathrm{K}$}},
  columns/T300K/.style={column name={$300\,\mathrm{K}$}},
  columns/T1000K/.style={column name={$1000\,\mathrm{K}$}},
  every head row/.style={before row=\toprule, after row=\midrule},
  every last row/.style={after row=\bottomrule},
]{gcsTable-constrelax.csv}

\vspace{2ex}

\textbf{Lognormal aggregates}

\vspace{0.5ex}

\begin{filecontents*}{gcsTable-lognormrelax.csv}
N,T3K,T10K,T30K,T100K,T300K,T1000K
30,0.434 ± 0.032,0.429 ± 0.039,0.440 ± 0.034,0.429 ± 0.038,0.427 ± 0.044,0.433 ± 0.038
100,1.040 ± 0.060,1.033 ± 0.059,1.032 ± 0.060,1.033 ± 0.056,1.007 ± 0.045,1.013 ± 0.075
300,2.311 ± 0.106,2.286 ± 0.103,2.253 ± 0.101,2.216 ± 0.080,2.192 ± 0.095,2.182 ± 0.056
\end{filecontents*}
\pgfplotstabletypeset[
  col sep=comma,
  string type,
  columns={N,T3K,T10K,T30K,T100K,T300K,T1000K},
  columns/N/.style={column name={$N$ (monomers)}},
  columns/T3K/.style={column name={$3\,\mathrm{K}$}},
  columns/T10K/.style={column name={$10\,\mathrm{K}$}},
  columns/T30K/.style={column name={$30\,\mathrm{K}$}},
  columns/T100K/.style={column name={$100\,\mathrm{K}$}},
  columns/T300K/.style={column name={$300\,\mathrm{K}$}},
  columns/T1000K/.style={column name={$1000\,\mathrm{K}$}},
  every head row/.style={before row=\toprule, after row=\midrule},
  every last row/.style={after row=\bottomrule},
]{gcsTable-lognormrelax.csv}

\end{table}

\end{document}